\documentclass[preprint,aps,pra]{revtex4}

\begin{document}

\voffset=0.5in

\title{Probabilistic Teleportation of a Qudit}

\author{Arun K. Pati} 
\email{akpati@iopb.res.in}
\author{Pankaj Agrawal}
\email{agrawal@iopb.res.in}

\affiliation{Institute of Physics, Sachivalaya Marg, Bhubaneswar-751005, Orissa, India}

\newcommand{\Nol}{{1 \over \sqrt{1 + |\ell|^ 2} }}
\newcommand{\Non}{{1 \over \sqrt{1 + |n|^2    } }}
\newcommand{\Nop}{{1 \over \sqrt{1 + |p|^2    } }}
\newcommand{\Nolp}{{1 \over \sqrt{1 + |\ell^{\prime}|^ 2} }}
\newcommand{\Nom}{{1 \over \sqrt{1 + |m|^2    } }}
\newcommand{\Nopp}{{1 \over \sqrt{1 + |p^{\prime}|^2    } }}

\newcommand{\be}{\begin{equation}}
\newcommand{\ee}{\end{equation}}
\newcommand{\bea}{\begin{eqnarray}}
\newcommand{\eea}{\end{eqnarray}}

\newcommand{\rag}{\rangle}
\def\ie{\hbox{\it i.e.}{}}
\def\eg{\hbox{\it e.g.}{}}

\date{\today}


\def\ra{\rangle}
\def\la{\langle}
\def\ver{\arrowvert}


\begin{abstract}
It is known that if the shared 
resource is a maximally entangled state then it
is possible to teleport an unknown state with unit fidelity and unit 
probability. However, if the shared resource is a non-maximally entangled 
state then one has to follow a probabilistic scheme where one can teleport 
a qubit with unit fidelity and non-unit probability. In this work, 
we investigate the feasibility of using partially entangled states as a 
resource for quantum teleportation of a qudit. 
We also give an expression for the probability of successful 
teleportation of an unknown qudit.
\end{abstract}


\maketitle






\section{Introduction}

    Quantum teleportation protocol plays an important role
   in the field of quantum computation and quantum communication.
   This protocol allows a sender to transmit an unknown quantum
   state to a receiver by using an entangled state as
   a quantum resource and by sending classical information via ordinary
channel \cite{bbcjpw}. Thus, the quantum teleportation protocol 
has become a paradigm example of quantum 
communication where the sender and the receiver are allowed to do local
   operations and classical communication (LOCC) only.
In the original protocol, the authors showed that if the sender
   (Alice) and the receiver (Bob) share a maximally entangled
   two-qubit state, then Alice can transmit an unknown qubit
   with unit probability and unit fidelity to Bob. They also 
   demonstrated how to teleport the 
   state of a qudit (a quantum system with $d$-dimensional Hilbert 
   space) with the help of a maximally entangled states of two qudits.  
   Subsequently, the teleportation protocol has been shown
   to work for a wide variety of bipartite systems including
   when entangled states are labeled by continuous parameters \cite{lv,bk}.
   The protocol has also been extended and shown to work when
   available quantum resource is a multi-partite entangled
   state \cite{ap1}. Experiments have also been done to demonstrate the
   feasibility of teleportation in laboratories \cite{db,dbe,af}.

   However in real life situations, it is most of the time not
   possible to have a maximally entangled state at one's disposal. 
Because of the interaction with the environment, the state of any system 
   would become a mixed state after a certain period. 
This problem of decoherence can be mitigated
   but cannot be completely overcome easily. 
Also, it may happen that the source does not produce 
perfect Einstein-Podolsky-Rosen (EPR) pairs rather non-maximally 
entangled pairs which is shared between Alice and Bob. 
Therefore, it is
   important to examine how the protocol would work with such
   resources. Especially, if we have non-maximally entangled state as a 
shared resource and we want to do quantum teleportation, 
then we have to pay some price. That is we have to compromise 
either in fidelity or in the success probability. If we are ready to pay 
the price for the success probability then it is possible to have unit fidelity
teleportation. And this scheme we call probabilistic quantum teleportation. 

It may be mentioned that in the literature, 
the possibility of teleportation 
   with both types of resources have been investigated. 
In the case of non-maximally entangled state as a resource,
   there are many possibilities. Some of these possibilities are
   entanglement concentration \cite{bbps}, use of non-maximally entangled 
basis \cite{ap,pa}, use of POVM instead of von Neumann measurement 
\cite{mor,son,roa}, use of higher dimensional entangled resource 
\cite{gour}, an antilinear operator description of the teleportation
\cite{kkaj}  and many more. 
   One of the possibilities is that of the probabilistic teleportation
   when one uses non-maximally entangled basis to make
   von Neumann measurement. This was discovered 
   in the case of qubits in Ref \cite{ap,pa}. This protocol
   has also been generalized to teleport $N$ qubits \cite{gordon}. It turns 
out  that when we have non-ideal EPR pair like general entangled state then it
is better to adopt our protocol. The probabilistic quantum teleportation scheme
is a single shot on demand protocol that allows perfect teleportation of a 
quantum system with unit fidelity but with a probability that is less than
unity. The main motive in this paper is to 
   extend this protocol to the case of higher dimensional 
quantum system like an unknown qudit.


      The paper is organized as follows. In Section II,
   we review the protocol for the case of maximally entangled
   quantum resource. In Section III, we present the probabilistic 
teleportation scheme for qubit and then generalize our results
   about the probabilistic teleportation to a qudit.
   Finally, in section IV, some conclusions are presented.

\section{Teleportation With Maximally Entangled State}

Before discussing the teleportation with non-maximally entangled states
as a quantum resource, we review the protocol for teleporting a qudit using
maximally entangled states as a quantum resource. Let us consider 
two observers Alice and Bob possessing qudits `1' and `2', respectively. 
Their qudits are in a maximally entangled state, which can be 
written as
\be
|\Phi \ra_{12} = {1 \over \sqrt{d}} \sum_{k=0}^{d-1}  |k \ra_{1} |k\ra_{2}.
\ee
Alice has a qudit `a' in the following unknown state
\be
|\psi\ra_{a} = \sum_{k=0}^{d-1} a_{k} |k \ra_{a}, 
\ee
where $a_{k}$ are unknown complex numbers. Alice wishes
to transmit this unknown state to Bob using local 
quantum operation and classical communication. To start with Alice 
can make a joint von Neumann measurement on her qudit `1' and the 
qudit `a'. To make this measurement she can use following Bell 
basis for qudits
\be
|\Psi_{\ell p}\ra  = {1 \over \sqrt{d}} 
\sum_{k=0}^{d-1}   e^{ 2 \pi i \ell k/d} |k \oplus p 
\ra | k\ra,
\ee
where $k \oplus p$ means sum of $k$ and $p$ modulo $d$. The indices 
  $\ell$ and $p$ can take integer values between $0$ and $d-1$.
Here, $|\Psi_{\ell p}\ra$ form a set of basis vectors for a two-qudit
system. 
We can invert this set of vectors and obtain
\be
|i j \ra  = {1 \over \sqrt{d}}
\sum_{\ell,m=0}^{d-1}  e^{ - 2 \pi i \ell k/d} 
\delta_{j,i\oplus m} |\Phi^{\ell m} \ra. 
\ee
Let us note that $|\Psi_{00}\ra \equiv  | \Phi \ra$. Now, we can rewrite 
the combined state of the system `a12' in terms of the above Bell basis
vectors for the system `a1' as follows
\be
|\psi \ra_{a} |\Phi \ra_{12} = {1 \over d} \sum_{\ell,n=0}^{d-1} 
|\Psi_{\ell n} \ra_{a1} U^{\dagger}_{\ell n}|\chi \ra_{2},
\ee
where these unitary operators $U_{nm}$ are given by
\be
U_{nm} = \sum_{k=0}^{d-1}   e^{ 2 \pi i nk/d} |k\ra \la k \oplus m|.
\ee
These unitary operators obey the following orthogonality condition
\be
{\rm Tr}(U^{\dagger}_{nm} U_{\ell p}) = d \; \delta_{n \ell} \; \delta_{mp}.
\ee
After Alice  makes the von-Neumann measurement in the Bell
  basis (3), she would obtain one of the possible $d^2$ results.
  She can convey the result of her measurement to Bob by sending 2 log$_{2}$d
  classical bits of information. After receiving this information Bob uses
  appropriate unitary operator $U_{nm} $ on his qudit to convert its state to
  that of the input state.  This completes the standard teleportation protocol.

   We note that here Alice succeeds in transmitting the state of the
  qudit `a' to Bob, irrespective of the result of her measurement.
  So the probability of the success is unity. After receiving the information
  from Alice, Bob can convert the state of his qudit `2' to that
  of the qudit `a' exactly. So the fidelity of the transmitted state
  is also unity. In other words, it is a case of perfect teleportation.
  However, as we shall see, when the shared entangled resource is not
  maximally entangled, one has to compromise either in probability or
  fidelity.

\section{Teleportation With Non-maximally Entangled State}

   We wish to now consider the situation, when available quantum resource
 is a non-maximally entangled two-qudit state. Earlier, a probabilistic 
teleportation scheme for qubit has been proposed \cite{ap}. Here, 
like the case of 
qubit, we examine the possibility of teleportation of a qudit when the shared 
state is a non-maximally entangled state. We will show that it is possible to
teleport a qudit with unit fidelity but with a probability that is less 
than unity.

\subsection{Probabilistic Teleportation of a Qubit}

For the sake of completeness we briefly review the probabilistic 
quantum teleportation of a qubit. Suppose Alice receives a qubit 
in an unknown state $|\psi \rag_a = \alpha |0\rag_a + \beta |1\rag_a$. 
She wishes to teleport this state to Bob using LOCC. However, here 
the pre-existing quantum channel is not a maximally entangled one,
but a pure non-maximally entangled state $|NME\ra$ which is given by
\be
|NME \rag_{12} = N \,(|00\rag_{12} + \, n |11\rangle_{12}),
\ee
where $n$ is a known complex number and $N=\Non$ is a real number. Alice and
Bob are in the possession of qubits $1$ and $2$, respectively. 
If Alice performs a measurement in the Bell basis on the system
`a1', then we know that the state 
$|\psi\rag_{a}$ cannot be teleported faithfully, \ie,
with unit fidelity and unit probability.
However, if a measurement is performed in a {\em non-maximally 
entangled} (NME)  basis having same amount of entanglement as
that of the shared resource then it is possible for Alice to 
teleport the state with unit fidelity, though not with unit 
probability \cite{ap}. 

To see this, let us introduce a set of basis vectors for 
two qubits possessed by Alice. Using the computational basis 
vectors $\{|00\rag, |01\rag,|10\rag, |11\rag \}$, we can 
define a set of mutually orthogonal NME basis vectors as follows
\bea
|\varphi^{+}_{\ell}\rag 
= L \,(|00\rag + \,\ell \,|11\rag) \nonumber\\
|\varphi^{-}_{\ell}\rag = 
L \, (\ell^{*}|00\rag - \,|11\rag) \nonumber\\
 |\psi^{+}_{p}\rag 
= P \, (|01\rag + \,p \,|10\rag)  \nonumber\\
|\psi^{-}_{p}\rag = P \,(p^{*}|01\rag - \,|10\rag)
\eea
Here $\ell$ and $p$ are complex numbers in general and 
$L = \Nol$ and $P = \Nop$ are real numbers.  
With the change of parameter $\ell$  and $ p$ values, 
this set interpolates between 
unentangled and maximally entangled set of basis vectors \cite{ap}. 


We can invert the transformations of (9) and use it to
rewrite the combined state of the input and resource systems as
\bea
|\psi\rag_{a} |NME \rag_{12} & = 
&  N \; (\alpha |0\rag_{a} + \, \beta |1\rag_{a})\,
(|00\rag_{12} +\, n |11\rag_{12})  \nonumber  \\
& = & N \; (\alpha |00\rag_{a1} |0\rag_{2} +
\, \alpha \,n |01\rag_{a1}|1\rag_{2} + \,\beta |10\rag_{a1} |0\rag_{2}
+\, \beta \,n |11\rag_{a1} |1\rag_{2} ) \nonumber \\
& = & N [|\varphi^{+}_{\ell}\rag_{a1} \ver f_1(\psi) \ra_2
+ |\varphi^{-}_{\ell} \rag_{a1}\ver f_2(\psi) \ra_2 
+ |\psi^{+}_{p}\rag_{a1} \ver f_3(\psi) \ra_2 
+ |\psi^{-}_{p}\rag_{a1} \ver f_4(\psi) \ra_2 ]
\eea
where $\ver f_1(\psi) \ra_2 = L ( \alpha |0\rag_{2} + n \beta \ell^{*} 
|1\rag_{2} ), 
\ver f_2(\psi) \ra_2 = L ( \ell \alpha |0\rag_{2} - n \beta |1\rag_{2}), 
\ver f_3(\psi) \ra_2 = P (\beta p^{*} |0\rag_{2} + \alpha  n |1\rag_{2})$, 
and $\ver f_4(\psi) \ra_2 = P (- \beta|0\rag_{2} + \alpha n p |1\rag_{2})$ 
are the unnormalized states. This expression
is the most general way of rewriting an unknown state and two qubit entangled
state. 

 As shown in Ref \cite{ap,pa},
if Alice makes the choice $ \ell = n = p^{*}$, or $\ell = n =
{1 \over p}$, or $\ell^{*} = {1 \over n} = p$, or $ \ell^{*} =
{1 \over n} = {1 \over p^{*}}$, then for any of these choices,
reliable teleportation is possible for only two out of four 
possible results of the measurement. For example, in the case of first
choice, when the outcome is $|\varphi^{-}_{\ell = n}\rag$, then the state at
Bob's hand will be $(\alpha |0\rag - \beta|1\rag)$ and when the outcome is
$|\psi^{+}_{p=n}\rag$, then the state at Bob's hand is
$(\beta |0\rag + \alpha|1\rag)$. Therefore, when Alice sends two classical
bits to Bob he will apply $\sigma_z$ in the former and $\sigma_x$ in the
later case to recover the unknown state with unit fidelity.
The total probability of this successful teleportation will be given by
\be
P_{\rm{succ}}= { 2 |n|^2 \over (1 + |n|^2)^2}.
\ee
An interesting observation in the case of qubit 
is that the above choice of parameters refers to the situation 
where the basis used for joint measurements and the resource 
state have the same amount of quantum entanglement, namely,
$E(|NME \rag)= (- \, N^2\rm{log}_{2}N^2 - N^2 \, |n|^2 \,
\rm{log}_{2}N^2 |n|^2)$.
Thus, we can say that using
$E = E(| NME \rag)= (- \, N^2\rm{log}_{2}N^2 - N^2 \, |n|^2 \,
\rm{log}_{2}N^2 |n|^2)$ amount of entanglement and two classical bits Alice
{\em can teleport an unknown state with unit fidelity and probability 
given in }(11). This is one of the main result discovered in \cite{ap}.

We would like to mention two important differences between 
the filtering approach and ours. In the
filtering approach one cannot proceed with the Bennett {\it et al}
protocol if the filtering does not succeed. Second, in filtering approach 
Alice needs to communicate three classical bits (one cbit at the stage of
filtering and two cbits after the Bell measurements). However, in our scheme
we can carry out the protocol given any non-maximally entangled 
state and it works just with two cbits. That is the reason our probability of 
success is lower than the filtering approach.



\subsection{Probabilistic Teleportation of a Qudit}

Let Alice and Bob share a non-maximally bipartite entangled state which 
is given in terms of the Schmidt decomposition form as
\be
|\Phi \ra_{12}  = \sum_{j=0}^{d-1} \sqrt{\lambda_{j}} |j \ra_{1} |j \ra_{2}, 
\ee
where $\lambda_{j}$'s are the Schimdt coefficients and $\sum_j \lambda_{j} =1$.
Alice and Bob have qudits `1' and `2', respectively in the above
entangled state. We wish to rewrite the above non-maximally entangled state 
for the sake of later convenience as follows
\be
|\Phi \ra_{12}  = D \sum_{j=0}^{d-1} d_{j} |j \ra_{1} |j \ra_{2}, 
\ee
where $ D = 1 /\sqrt{ \sum_{j=0}^{d-1} |d_{j}|^2} $ is the normalization
constant and $\sqrt{\lambda_j } = D d_j$. Alice has a qudit 
`a' in the unknown state
\be
|\psi \ra_{a} = \sum_{k=0}^{d-1} a_{k} |k\ra_{a},
\ee
where $a_{k}$ are unknown complex coefficients. Alice now wishes
to transmit this state to Bob.

Lt us introduce a set of general two-qudit entangled states as follows
\be
|\Phi^{\ell m}\ra  =
 N_{\ell m} \sum_{j=0}^{d-1} c^{\ell m}_{j} |j \ra |j \oplus m \ra,
\ee
where $N_{\ell m} $ is the normalization constant and is equal to
$1 /\sqrt{ \sum_{j=0}^{d-1} |c^{\ell m}_{j}|^2} $.
If the above is part of a set of $d^2$ orthonormal basis vectors,
then the coefficients $c^{\ell m}_{j}$ should satisfy the following 
condition, as the states $|\Phi_{\ell m} \ra$ would
be orthonormal
\be
N_{\ell m} N_{pm} \sum_{k=0}^{d-1} c^{* \ell m}_{k} c^{p m}_{k} 
= \delta^{\ell p},
\ee
where the indices $\ell, p, m$ and $n$ take integer values
   between $0$ and $d-1$.
The set of vectors $\{ |\Phi^{\ell m}\ra \}$ will be used as a 
measurement basis (when appropriate condition is satisfied).

We can invert the equation (16) and obtain
\be
|i j \ra  =
\sum_{\ell,m=0}^{d-1} N_{\ell m} c^{* \ell }_{j} \delta_{j,i\oplus m} |\Phi^{\ell m} \ra 
\ee
Let us rewrite the state of the combined of the system `a12' as
\bea
|\psi \ra_{a} |\Phi \rag_{12} & = & D \sum_{n,j=0}^{d-1} a_{n} d_{j} |n \ra_{a}
                    |jj \ra_{12} \nonumber  \\
             & = & D \sum_{n,j=0}^{d-1} a_{n} d_{j} |nj \ra_{a1} |j \ra_{2}
  \nonumber  \\
     & = & D \sum_{n,\ell,m=0}^{d-1} N_{\ell m} a_{n} d_{n \oplus m} 
c^{* \ell m}_{n} |\Phi^{\ell m} \ra |n \oplus m \ra \nonumber\\
&=& D  \sum_{\ell,m=0}^{d-1}  N_{\ell m} |\Phi^{\ell m} \ra 
|f_{\ell m}(\psi) \ra, 
\eea
where $|f_{\ell m}(\psi) \ra = \sum_n  a_{n} d_{n \oplus m} c^{* \ell m}_{n}
|n \oplus m \ra $ is a set of unnormalized kets. Note that 
$|f_{\ell m}(\psi) \ra$ has information about the unknown state.
For the quantum teleportation process to succeed, on the right hand side 
we should have the kets $|f_{\ell m}(\psi) \ra$ proportional to 
the unknown state upto local unitary transformations, i.e., 
\be
U^{\dagger}_{nm} |\psi \ra = \sum_{\ell=0}^{d-1} a_{\ell} \, 
e^{- 2 \pi i n \ell/d} |\ell \oplus m \ra.
\ee
If it is so, then after receiving classical information which is a function of
$(n m)$ Bob can apply $U_{nm}$ to his qudit and convert its state to that 
of $|\psi \ra $. That will complete the probabilistic quantum teleportation 
protocol for an unknown qudit.

From the last two equations, we notice that the condition for quantum 
teleportation to succeed with a finite probability is given by 
\be
       c^{ \ell m}_{n} = {1 \over d^{*}_{n \oplus m} } e^{ 2 \pi i \ell n/d}
                       = f_{n \oplus m} e^{ 2 \pi i \ell n/d}, 
\ee
where $f_{n \oplus m}$ is a complex number with the magnitude and phase
which are inverse of $d^{*}_{n \oplus m} $. As said before, if we could 
do this then we have
\be
|\psi \ra_{a} |\Phi \rag_{12}  =  D \sum_{\ell,m=0}^{d-1} N_{\ell m} 
             |\Phi^{\ell m} \ra_{a1} U^{\dagger}_{\ell m} |\psi \ra_{2}.
\ee
Then a von-Neumann measurement in the basis $\{|\Phi^{\ell m} \ra \}$
  can lead to successful teleportation with unit fidelity and 
the probability is $|D\, N_{\ell m}|^2$
  for an outcome $|\Phi^{\ell m} \ra $. However, as in the qubit case, not
  all measurements would lead to successful teleportation. This will happen
  only in some of the cases. To find the number of such cases, we note that
  the coefficients $c^{ \ell m}_{n} $ must satisfy the orthonormality
  condition (17). The successful teleportation requirement (20) may not 
  always satisfy (17). Combining the two conditions we get

\be
        \sum_{n=0}^{d-1} ({1 \over |d_{n \oplus m}|^2}) \;\;
        {1 \over \sum_{p=0}^{d-1} ({1 \over |d_{p \oplus m}|^2})} \;\;
         e^{-2 \pi i (\ell - k) n/d} = \delta_{\ell k}.
\ee
This condition can only be satisfied when $\ell = k$. So the teleportation
   is successful only $d$ out of $d^2$ times. 
We can understand this result as follows. For a system of two qudits,
  these vectors $\{|\Phi^{\ell m} \ra \}$ naturally falls into
  $d$ classes. Each class is labeled by $\ell$. Within each class,
  there are $d$ states, which are labeled by $m$. These $d$ classes
  are orthogonal to each other. With the choice (20) for the
  coefficients $c^{\ell m}_{n}$, teleportation is successful 
  once for each class.

We can calculate explicitly the total probability of success in teleporting 
an unknown qudit. This is given by 
\be
         P_{\rm succ} = {d \over \sum_{n=0}^{d-1} |d_{n}|^2} \;\;
        {1 \over \sum_{k=0}^{d-1} ({1 \over |d_{k}|^2})} = 
{d \over \sum_{k=0}^{d-1} {1 \over \lambda_k}}
\ee
Thus, we can say that using $E(|\Phi\ra) = 
- \sum_n \lambda_n \log_2 \lambda_n $ amount of entanglement and $2 \log_2 d$ 
number of classical bits one can teleport an unknown qudit with unit fidelity 
but with a probability $P_{\rm succ}$ that is less than unity.
We can check that this result reduces to the results for the qubit
  case. For qubit case $d = 2$. So one can succeed twice. This is in accord
  with the result of Ref \cite{ap}. As a consistent check, 
if we substitute appropriately
  for the values of $d_{n}$, then the above expression for the success
  probability also reduces to that of the qubit case, i.e., $ P_{\rm succ} =
{2 |n|^2 \over (1 + |n|^2)^2 }$. Another remark is the following: In the case 
of probabilistic teleportation of a qubit, it was observed that 
the non-maximally entangled measurement basis had the same entanglement as
the shared resource. However, in the case of qudit, the non-maximally 
entangled measurement basis do not have same amount of entanglement as the shared
resource. Because, the entanglement of $\{|\Phi^{\ell m} \ra \}$ is 
$E(|\Phi^{\ell m} \ra ) =  -\sum_n N_{\ell m}^2 |c_{\ell m}|^2 \log_2  
 N_{\ell m}^2 |c_{\ell m}|^2 $ which in general cannot be same as 
$E(|\Phi\ra) = - \sum_n \lambda_n \log_2 \lambda_n $. Only, in the case of 
$d=2$ they coincide for the teleportation condition (20).

Furthermore, we can say that one can amplify the probability 
statistically by repetitions.
We know that the reciprocal of the {\em average} success probability
must be the number of repetitions $R$ that are required in order to
successfully (all the time) teleport an unknown state with unit fidelity.
We see that one shall need on the average 
at least $R= 1/P_{\rm succ}$ repetitions to get a
faithful teleportation with unit probability. Therefore, if Alice and Bob share
$R E(|\Phi\ra) $ pairs of non-maximally entangled state they can
successfully teleport an arbitrary qudit state using local operation 
and $2 R \log_2 d $ bits
of classical communication. 
One can see that as the degree of entanglement
increases, the number of required repetitions decreases and becomes one
for maximally entangled states as expected. It becomes infinite
for the untangled resource state. Therefore, if Alice and Bob do not have 
prior shared entanglement then it will be impossible to teleport an unknown
state with unit fidelity.

\section{Conclusions}

In this paper we have investigated how to teleport an unknown quantum state 
when Alice and Bob have shared a general bipartite pure entangled state in
$d \times d$. 
Obviously, one cannot teleport the state with unit fidelity and unit 
probability. But if we pay the price for the success probability then it is 
possible to do quantum teleportation with unit fidelity. This we call 
the probabilistic quantum teleportation 
scheme. When the available quantum resource is not a maximally entangled state,
then it is advisable to implement our scheme which is a single shot, on demand 
teleportation protocol without having recourse to quantum filtering or 
entanglement concentration. Inspired by the scheme for qubit, we 
have examined the possibility of teleporting the
    state of an unknown qudit using non-maximally entangled state as
    a quantum resource. We find that quantum teleportation is
    possible again only probabilistically, i.e., we can indeed teleport an 
unknown qudit with unit fidelity but with a probability less than unity.
It is found that only $d$ times out of $d^2$ measurements, the state could 
be teleported with unit fidelity. We have given an expression for the 
success probability. We hope that with current technology one should be able 
to implement the probabilistic quantum teleportation protocol for a 
qubit and a qudit in near future. Also, one may investigate how to generalize
the probabilistic teleportation protocol for continuous variable systems which 
seems to be a non-trivial task.

\end{document}